\documentclass[aps,pre,10pt,twocolumn,superscriptaddress,showpacs]{revtex4-1}
\usepackage{graphicx}
\usepackage{color}
\usepackage{bm}

\begin{document}

\title{Sparse representation for potential energy surface}
\author{Atsuto \surname{Seko}}
\email{seko@cms.mtl.kyoto-u.ac.jp}
\affiliation{Department of Materials Science and Engineering, Kyoto University, Kyoto 606-8501, Japan}
\affiliation{Center for Elements Strategy Initiative for Structure Materials (ESISM), Kyoto University, Kyoto 606-8501, Japan}
\author{Akira \surname{Takahashi}}
\affiliation{Department of Materials Science and Engineering, Kyoto University, Kyoto 606-8501, Japan}
\author{Isao \surname{Tanaka}}
\affiliation{Department of Materials Science and Engineering, Kyoto University, Kyoto 606-8501, Japan}
\affiliation{Center for Elements Strategy Initiative for Structure Materials (ESISM), Kyoto University, Kyoto 606-8501, Japan}
\affiliation{Nanostructures Research Laboratory, Japan Fine Ceramics Center, Nagoya 456-8587, Japan}

\date{\today}

\begin{abstract}
We propose a simple scheme to estimate the potential energy surface (PES) for which the accuracy can be easily controlled and improved. 
It is based on model selection within the framework of linear regression using the least absolute shrinkage and selection operator (LASSO) technique.
Basis functions are selected from a systematic large set of candidate functions.
The sparsity of the PES significantly reduces the computational cost of evaluating the energy and force in molecular dynamics simulations without losing accuracy.
The usefulness of the scheme for describing the elemental metals Na and Mg is clearly demonstrated.
\end{abstract}

\maketitle
\section{Introduction}
About sixty years ago, molecular dynamics (MD) was proposed as a tool to model a collection of interacting atoms within classical mechanics\cite{alder1959studies}.
The accuracy of the potential energy surface (PES) in terms of the atomic positions is crucial in performing MD simulations.
The PES determines the forces acting on atoms that originate from atomic interactions and thereby the motion of atoms.
A reliable PES can be obtained by directly computing the energy and forces acting on atoms for an atomic configuration at each time step by density functional theory (DFT) calculation\cite{PhysRev.136.B864,PhysRev.140.A1133}, known as first-principles MD calculation\cite{car1985unified}.
Owing to advances in computational power and algorithms, the first-principles MD calculation can be carried out for large systems composed of thousands of atoms\cite{PhysRevLett.94.158103}.
Also, computational approaches for efficiently performing the first-principles MD simulation have been proposed (e.g., \cite{PhysRevLett.98.066401}). 
However, because the first-principles MD calculation is still computationally demanding, many different interatomic potentials have often been used instead, particularly for considerably large systems, including the Lennard-Jones, Morse, embedded atom method (EAM) and Tersoff potentials\cite{lennard1924determination,PhysRev.34.57,tersoff1986new,torrens1972interatomic,carlsson1990beyond,finnis2003interatomic}.
Parameters in the potential functions are optimized by fitting to a set of experimental data.
First-principles results have also been used to derive more accurate interatomic potentials (e.g., \cite{Sbraccia2002147,doi:10.1021/jp412808m}). 

Recently, an alternative approach to estimating the PES from a large set of DFT energies was demonstrated\cite{behler2007generalized,bartok2010gaussian}.
In this approach, the information of atomic positions in a structure is transformed into various descriptors\cite{behler2011atom,jose2012construction,bartok2013representing}, although it is not clear which descriptors should be used for the target material.
Then, nonlinear regression techniques such as neural networks\cite{behler2007generalized} and the Gaussian process\cite{bartok2010gaussian} have been used to bridge the energy and descriptors because a complex function of the PES in terms of the descriptors must be estimated without a priori knowledge of the target material.
Using these techniques, the accuracy of the PES is generally much better than that obtained using conventional interatomic potentials.
Another advantage is their applicability to a wide range of materials including metallic\cite{PhysRevB.81.184107,PhysRevB.85.045439}, covalent\cite{behler2007generalized,bartok2010gaussian,PhysRevB.85.174103} and ionic materials\cite{PhysRevB.83.153101}.

In this study, we show another simple procedure for estimating the PES from a large set of DFT calculations.
We here propose to use the framework of linear regression even though the PES is a complex function in terms of the atomic positions.
Compared with nonlinear regression techniques, linear regression has a number of advantages as follows: 1) Accuracy can be controlled in a transparent manner.
2) Regression coefficients are generally determined quickly using a standard least-squares technique.
3) The number of regression coefficients does not explicitly depend on the size of the input data set.
As descriptors for expressing atomic positions, a systematic set of simple basis functions is adopted.
By using a combination of linear regression and systematic basis functions, the accuracy of the PES can be easily controlled and improved.

We use two methods for the estimation of the PES for the elemental metals Na and Mg.
One is the linear ridge regression technique\cite{hastieelements} using all the basis functions in the given set of basis functions.
The other uses the least absolute shrinkage and selection operator (LASSO) technique\cite{tibshirani1996regression} to perform the automatic selection of basis functions.
The LASSO enables us to obtain a well-optimized sparse expression for the PES with a small number of nonzero coefficients.

\section{Linear model for potential energy surface}
\begin{figure}[tbp]
\begin{center}
\includegraphics[width=\linewidth,clip]{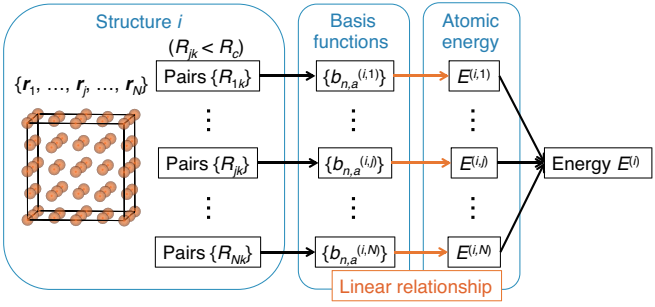} 
\caption{
Relationship between atomic positions and total energy in the linear model for the PES of structure $i$.
It is assumed that the energy of atom $j$, $E^{(i,j)}$, and the set of basis functions for atom $j$, $\{ b_{n,a} ^{(i,j)} \}$, have a linear relationship.
The set of basis functions is here calculated from the set of pair distances $\{ R_{jk}^{(i)} \}$ smaller than the cutoff radius $R_c$, which are obtained from the positions of all atoms.
}
\label{linear:linear_model}
\end{center}
\end{figure}

A linear model for describing the energy of a structure is shown in Fig. \ref{linear:linear_model}.
This model is invariant to the translation and exchange of atoms and is based on the widely accepted idea that the total energy of a structure is equal to the sum of its atomic energies\cite{behler2007generalized,bartok2010gaussian}.
The total energy of structure $i$ is expressed as
\begin{math}
E^{(i)} = \sum_{j} E^{(i,j)},
\end{math}
where $E^{(i,j)}$ denotes the contribution of atom $j$ to the total energy of structure $i$.
Then, a linear relationship between the atomic energy and $M$ basis functions is introduced as
\begin{equation}
E^{(i,j)} = \bm{w}^\top \bm{b}^{(i,j)},
\end{equation}
where $\bm{w} = [w_1, \cdots, w_M]^\top$ and $\bm{b}^{(i,j)} = [b_1^{(i,j)}, \cdots, b_M^{(i,j)}]^\top$ denote the regression coefficients and basis functions for atom $j$ of structure $i$, respectively.
By applying the same regression coefficients to identical atomic species, the total energy is derived as
\begin{equation}
\label{linear:total_energy_basis}
E^{(i)} = \bm{w}^\top \bm{c}^{(i)},
\end{equation}
where the $M$-vector $\bm{c}^{(i)}$ satisfies the equation 
\begin{math}
\bm{c}^{(i)} = \sum_{j} \bm{b}^{(i,j)} .
\end{math}

Here, functions with a simple form are implemented as basis functions.
The basis function which is the $a$th power of the $n$th element, $ b_{n,a}^{(i,j)}$, is written as
\begin{equation}
b_{n,a}^{(i,j)} = \left[\sum_k f_n(R_{jk}^{(i)}) \cdot f_c(R_{jk}^{(i)})\right]^a, 
\label{linear:basis_function}
\end{equation}
where $a$ is a positive integer and $R_{jk}^{(i)}$ denotes the distance between atoms $j$ and $k$ of structure $i$.
The sum is taken over all atoms within a cutoff radius $R_c$.
For $f_n(R_{jk}^{(i)})$, we adopted various types of systematic and analytical pairwise functions, which are Gaussian, cosine, Bessel, Neumann, polynomial and Gaussian-type orbital (GTO) functions.
$f_c(R_{jk}^{(i)})$ is a smooth pairwise cutoff function which is exactly zero at a distance greater than the cutoff radius $R_c$.
We use a cosine-based cutoff function as used in Ref. \onlinecite{behler2007generalized}.
Since the product of $f_n(R_{jk}^{(i)})$ and $f_c(R_{jk}^{(i)})$ is pairwise, an exponential form of the sum of the pairwise functions is introduced to take many-body effects into account.
Note that the use of a pairwise potential causes well-known serious problems for the description of the PES\cite{carlsson1990beyond,finnis2003interatomic}. 
For instance, pairwise interatomic potentials satisfy the Cauchy relationship for the elastic constants of $C_{12} = C_{44}$ in fcc crystals; this is an artifact. 
The cohesive energy is forced to exhibit a linear dependence on the coordination number.

\section{Estimation of potential energy surface}
\subsection{DFT calculations}
The PESs for the elemental metals Na and Mg described by the linear model were estimated from a large set of DFT calculations using the linear regression techniques.
As training data for the regressions, 1600 configurations were prepared on the basis of face-centered-cubic (fcc), body-centered-cubic (bcc), hexagonal-closed-packed (hcp) and simple cubic (sc) structures for both Na and Mg. 
They were generated by random distortions of the ideal fcc, bcc, hcp and sc structures, in which the atomic positions and lattice constants were fully optimized.
In addition to the training data, 400 configurations were prepared by the same procedure as training data to examine the predictive power for structures that were not included in the training data.
DFT calculations were performed for a total of 2000 configurations for both Na and Mg using the plane-wave basis projector augmented wave (PAW) method\cite{PAW1,PAW2} within the Perdew-Burke-Ernzerhof exchange-correlation functional\cite{GGA:PBE96} as implemented in the \textsc{vasp} code\cite{VASP1,VASP2}.
The total energies converged to less than 10$^{-2}$ meV/supercell.
For the ideal structures, the atomic positions and lattice constants were optimized until the residual forces became less than $10^{-3}$ eV$/$\AA.


\subsection{Linear ridge regression}
Subsequently, we constructed PESs for Na and Mg from the training data using linear ridge regression\cite{hastieelements}.
Linear ridge regression is one of the shrinkage methods and shrinks the regression coefficients by imposing a penalty.
When matrix $\bm{X} = [\bm{c}^{(1)}, \cdots, \bm{c}^{(N)}]^\top $ is composed of basis functions of the training data, the ridge coefficients minimize a penalized residual sum of squares expressed as
\begin{equation}
||\bm{X}\bm{w} - \bm{y}||_2^2 + \lambda ||\bm{w}||_2^2 , 
\label{linear:ridge_equation}
\end{equation}
where $\bm{y}$ denotes the DFT energies of the training data and $||\cdot||_2$ is the $L_2$-norm.
This is referred to as $L_2$ regularization. 
The regularization coefficient $\lambda$ controls the magnitude of the penalty.
The solution is given only in terms of matrix operations as 
\begin{math}
\bm{w} = (\bm{X}^\top \bm{X} + \lambda \bm{I})^{-1} \bm{X}^\top \bm{y},
\end{math}
where $\bm{I}$ denotes the unit matrix.
When the regression coefficients of many correlated variables in a linear regression model are determined without including the penalty term, they can be poorly determined and exhibit a large variance.
The penalty on the coefficients alleviates this problem.

\begin{table}[tbp]
\caption{
RMSEs for the test data of PESs obtained by linear ridge regression using 240 cosine basis functions,
\begin{math}
f_n(R) = \cos (\beta R),
\end{math}
with $R_c=7.0$ and $\lambda = 0.001$.
The interval and minimum value of $\beta$ are fixed to 0.05, and the maximum values of $\beta$ and $a$ control the number of basis functions.
}
\label{linear:ridge_cosine_rmse}
\begin{ruledtabular}
\begin{tabular}{ccc}
& Na (meV/atom) & Mg (meV/atom)\\
\hline
Cosine ($a_{\rm max} = 1$) & 7.3 & 11.8 \\
Cosine ($a_{\rm max} = 2$) & 1.6 & 2.6 \\
Cosine ($a_{\rm max} = 3$) & 1.4 & 1.6 
\end{tabular}
\end{ruledtabular}
\end{table}

For each PES constructed using a set of basis functions, we calculated the root-mean-square error (RMSE) between energies for the test data predicted by the DFT calculation and those predicted using the PES; this can be regarded as the predictive power of the PES.
Table \ref{linear:ridge_cosine_rmse} shows the RMSEs of the PESs for Na and Mg constructed from 240 systematic cosine basis functions, where the RMSE converges with increasing number of basis functions.
PESs constructed from basis functions with $a=1$, in which only pairwise interactions are considered, have low predictive powers for both Na and Mg.
On the other hand, increasing the maximum value of $a$, $a_{\rm max}$, improves the predictive power of PESs substantially.
Using 240 cosine basis functions with $a_{\rm max} = 3$ and $R_c = 7.0$ \AA, the RMSEs for Na and Mg are 1.4 and 1.6 meV/atom, respectively.
By increasing the cutoff radius to $R_c = 9.0$ \AA, the RMSE reaches a considerably small value of 0.4 meV/atom for Na, whereas it remains almost unchanged for Mg.

\begin{figure}[tbp]
\begin{center}
\includegraphics[width=0.99\linewidth,clip]{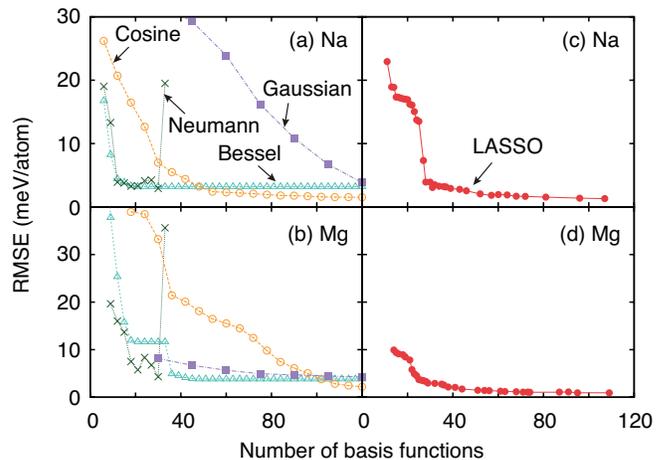} 
\caption{
RMSEs for the test data of the PESs constructed by linear ridge regression using various types of basis functions with $a_{\rm max} = 3$, $R_c=7.0$ and $\lambda = 0.001$ (a) for Na and (b) for Mg.
RMSEs optimized by the LASSO are also shown (c) for Na and (d) for Mg.
For the Bessel and Neumann basis sets, the number of basis functions is controlled by the maximum order of the basis functions.
For the Gaussian basis set, the number of basis functions is controlled by the maximum values of the internal parameters of the Gaussian.
}
\label{linear:LASSO_rmse}
\end{center}
\end{figure}

Then, we employed other types of single systematic basis functions with $a_{\rm max} = 3$, i.e., Gaussian, Bessel and Neumann basis functions, in addition to cosine basis functions.
Figures \ref{linear:LASSO_rmse} (a) and (b) show the dependence of the RMSE on the number of basis functions for Na and Mg, respectively.
When a small number of basis functions $(<20)$ are used, the Neumann basis set is the best among the four basis sets for both Na and Mg, although the RMSE is rather large.
On the other hand, when the number of basis functions is increased, the cosine basis set is satisfactory for increasing the accuracy for both Na and Mg, although a sufficient number of basis functions are needed for the convergence of the RMSE for Mg.

After applying the single basis functions to the estimation of the PES, we considered all combinations $(= 2^4 - 1 =15)$ of the four basis sets with $a_{\rm max} = 3$.
The Gaussian and cosine basis sets were composed of 120 basis functions.
The Bessel and Neumann basis sets were composed of 60 and 30 basis functions, respectively.
It was found that no combinations improve the RMSE for Na, whereas the combination of Gaussian, cosine and Bessel basis sets gives the best prediction for Mg with an RMSE of 0.9 meV/atom.

\subsection{LASSO}
The results of linear ridge regression indicate that the energy can be expressed by a linear relationship with simple basis functions depending only on the distances between atoms.
However, it appears that this method leads to the use of a large number of unnecessary basis functions to describe the PES.
To avoid this problem, the LASSO technique\cite{tibshirani1996regression,hastieelements} is applied, which enables us not only to provide a solution for linear regression but also to obtain a sparse representation with a small number of nonzero regression coefficients.
The LASSO is another shrinkage method, similar to ridge regression.
Using a large set of candidates composed of various types of systematic basis functions for the LASSO, three types of unknown features can be simultaneously optimized, i.e., the type of basis functions, the internal parameters of the basis functions and the number of basis functions.

The LASSO minimization function is defined as
\begin{equation}
||\bm{X}\bm{w} - \bm{y}||_2^2 + \lambda ||\bm{w}||_1, 
\end{equation}
where $||\cdot||_1$ denotes the $L_1$-norm.
The $L_2$ ridge penalty in Eqn. (\ref{linear:ridge_equation}) is replaced by the $L_1$ LASSO penalty.
The parameter $\lambda$ controls the trade-off relationship between sparsity and accuracy.
The LASSO solution is computed using a general quadratic programming technique, for which efficient algorithms are available.

The candidate basis functions were composed of a large number of Gaussian, cosine, Bessel, Neumann, polynomial and GTO basis functions, generated with fine intervals for the internal parameters of the basis functions.
The total number of candidate basis functions was 8455, which was much larger than the number of structures in the training data.
Sparse representations were then extracted from the set of candidate basis functions by the LASSO.
Figures \ref{linear:LASSO_rmse} (c) and (d) show the RMSEs of PESs obtained by the LASSO for Na and Mg, respectively.
The RMSE of the LASSO PES decreases more rapidly than those of PESs constructed from the single basis functions.
In other words, the LASSO PES requires a much smaller number of basis functions than linear ridge regression. 
For Na, a sparse representation with an RMSE of 1.3 meV/atom was obtained only using 107 basis functions, which is almost the same accuracy as the PES constructed from 240 cosine basis functions with an RMSE of 1.4 meV/atom.
Moreover, the LASSO PES has about 18 times fewer regression coefficients than a neural network potential with 1901 coefficients\cite{PhysRevB.81.184107}, although the RMSE of the LASSO PES is slightly larger than that of the neural network potential of 0.91 meV/atom.

\begin{figure}[tbp]
\begin{center}
\includegraphics[width=0.7\linewidth,clip]{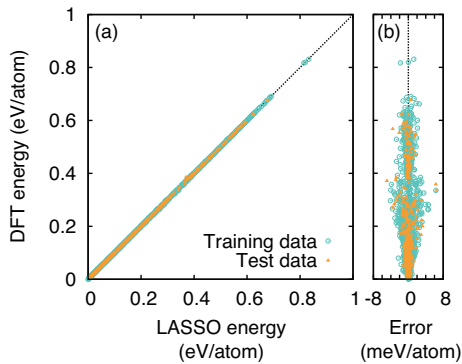} 
\caption{
(a) Energies predicted by the LASSO PES and DFT for Mg, measured from the energy of the ideal hcp structure.
(b) Differences between energies predicted by the LASSO PES and DFT.
}
\label{linear:error-Mg}
\end{center}
\end{figure}

It is apparent that the LASSO is more advantageous for Mg than for Na.
The obtained sparse representation with 95 basis functions for Mg has an RMSE of 0.9 meV/atom, which is almost half of that for the PES constructed from 240 cosine basis functions.
Figure \ref{linear:error-Mg} shows a comparison of the energies predicted by the LASSO PES and DFT for Mg.
As can be seen in Fig. \ref{linear:error-Mg}, there is little difference between the prediction errors for the training and test data.
In addition, a dependence of the prediction error on the energy is not clearly observed despite the wide range of structures included in both the training and test data.

\begin{figure}[tbp]
\begin{center}
\includegraphics[width=\linewidth,clip]{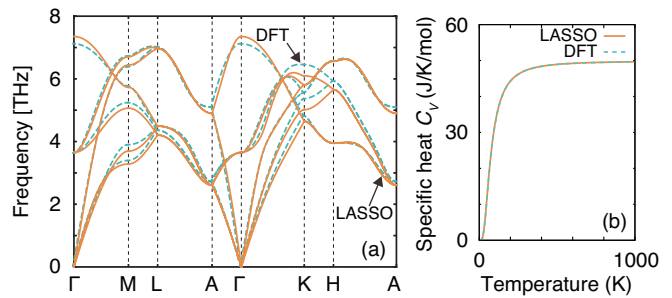} 
\caption{
(a) Phonon dispersion and (b) specific heat at a constant volume for hcp Mg obtained from the LASSO PES (solid lines) compared with those obtained by the DFT calculation (dotted lines).
To evaluate the dynamical matrix, each atomic position is displaced by 0.01 \AA.
}
\label{linear:phonon-Mg}
\end{center}
\end{figure}

Once the PES is constructed within the linear model, the forces acting on atoms can be analytically computed using Eqns. (\ref{linear:total_energy_basis}) and (\ref{linear:basis_function}).
Here the accuracy of the LASSO PES for the force calculation can be examined by comparing phonon dispersions computed by the LASSO PES and DFT.
The phonon dispersions and related thermodynamic properties were calculated by the supercell approach\cite{parlinski1997first}.
The phonon calculations were performed using the \textsc{phonopy} code\cite{PhysRevB.78.134106}.
Figures \ref{linear:phonon-Mg} (a) and (b) show the phonon dispersion and specific heat at a constant volume for hcp Mg, respectively, using the LASSO PES and DFT calculation.
Since the phonon dispersion of the LASSO PES does not differ from that calculated by DFT, the specific heats are in good agreement.
This demonstrates that the LASSO PES is sufficiently accurate to perform atomistic simulations of solids with similar accuracy to the DFT calculation.

\section{Conclusion}
We have introduced a simple scheme to estimate the PES for which the accuracy can be easily controlled and improved. 
We have applied it to describe the elemental metals Na and Mg. 
We found that the energy can be expressed by a linear relationship with simple basis functions depending only on distances between atoms. 
Using the LASSO, a sparse set of meaningful basis functions for expressing the PES can be easily extracted from a large number of candidate functions. 
As a result, we have obtained a sparse PES with prediction errors of 1.3 and 0.9 meV/atom for Na and Mg, respectively. 
The present method can accelerate to increase the accuracy of atomistic simulations while decreasing the computational costs.

\begin{acknowledgments}
This study was supported by a Grant-in-Aid for Scientific Research (A) and a Grant-in-Aid for Scientific Research on Innovative Areas ``Nano Informatics'' (grant number 25106005) from Japan Society for the Promotion of Science (JSPS).
\end{acknowledgments}

\bibliography{linear}

\end{document}